\documentclass[aps,prab,reprint,superscriptaddress]{revtex4-2}

\usepackage{comment}
\usepackage{graphicx}
\usepackage[percent]{overpic}
\usepackage{booktabs}
\usepackage{siunitx}
\usepackage{amsmath}

\usepackage{xcolor}
\usepackage[normalem]{ulem}
\usepackage{placeins}

\newif\ifshowchanges
\showchangesfalse

\ifshowchanges
  
  \newcommand{\remove}[1]{\textcolor{black}{\sout{#1}}}
\else
  
  \newcommand{\remove}[1]{}
\fi

\newcommand{\JAEA}{Japan Atomic Energy Agency (JAEA), Tokai, Naka, Ibaraki 319-1195, Japan}
\newcommand{\KEK}{High Energy Accelerator Research Organization, Ibaraki 319-1106, Japan}
\newcommand{\IwateUniv}{Graduate School of Science and Engineering, Iwate University, Morioka, Iwate 020-8551, Japan.}
\newcommand{\OkayamaUniv}{Research Institute for Interdisciplinary Science, Okayama University, Okayama 700-8530, Japan}
\newcommand{\NagoyaUniv}{Graduate School of Science, Nagoya University, Nagoya, Aichi 464-8602, Japan}
\newcommand{\KMI}{Kobayashi-Maskawa Institute for the Origin of Particles and the Universe, Nagoya University, Nagoya, Aichi 464-8602, Japan}
\newcommand{\UnivofTokyo}{Graduate School of Science, University of Tokyo, 7-3-1 Hongo, Bunkyo-ku, Tokyo 113-0033, Japan}

\begin{document}

\title{A diagnostic system of 5.7~keV muon beam for muon accelerator}

\author{M.~Wada} \email{mayuwada2001@g.ecc.u-tokyo.ac.jp} \affiliation{\UnivofTokyo}
\author{M.~Kimura} \email{masato.kimura@phys.s.u-tokyo.ac.jp} \affiliation{\UnivofTokyo}

\author{S.~Aritome} \affiliation{\UnivofTokyo}
\author{E.~Cicek} \affiliation{\KEK}
\author{K.~Futatsukawa} \affiliation{\KEK}
\author{K.~Hirai} \affiliation{\NagoyaUniv}
\author{M.~Hoshiai} \affiliation{\NagoyaUniv}
\author{T.~Iijima} \affiliation{\NagoyaUniv}\affiliation{\KMI}
\author{Y.~Imai} \affiliation{\OkayamaUniv}
\author{K.~Inami} \affiliation{\NagoyaUniv}\affiliation{\KMI}
\author{K.~Ishida} \affiliation{\KEK}
\author{S.~Kamioka} \affiliation{\KEK}
\author{Y.~Kawase} \affiliation{\IwateUniv}
\author{A.~Kondo} \affiliation{\NagoyaUniv}
\author{Y.~Kondo} \affiliation{\JAEA}
\author{M.~Lyu} \affiliation{\UnivofTokyo}
\author{T.~Mibe} \affiliation{\UnivofTokyo}\affiliation{\KEK}
\author{Y.~Nagatani} \affiliation{\KEK}
\author{Y.~Nakazawa} \affiliation{\KEK}
\author{S.~Ogawa} \affiliation{\KEK}
\author{Y.~Oishi} \affiliation{\KEK}
\author{M.~Otani} \affiliation{\KEK}
\author{N.~Saito} \affiliation{\UnivofTokyo}\affiliation{\KEK}
\author{K.~Shimomura} \affiliation{\KEK}
\author{K.~Suzuki} \affiliation{\KMI}
\author{T.~Takayanagi} \affiliation{\JAEA}
\author{K.~Ueda} \affiliation{\NagoyaUniv}
\author{S.~Uetake} \affiliation{\OkayamaUniv}
\author{S.~Yamamoto} \affiliation{\OkayamaUniv}
\author{T.~Yamazaki} \affiliation{\KEK}
\author{M.~Yang} \affiliation{\KEK}

\date{\today}

\begin{abstract}
Realization of a low-emittance muon beam through the acceleration of \si{\kilo\electronvolt}-scale muons requires the injection of a suitably matched beam into an accelerator, since beam mismatch can lead to emittance growth and reduced acceleration efficiency.
In one such scheme, muons are first thermalized to
room temperature and then injected into a linear
accelerator. Non-destructive diagnostics are challenging because of the low energy and low intensity.
We developed a compact low-energy muon diagnostic system compatible with the accelerator under construction at J-PARC.
The system is designed to evaluate beam conditions required for precise tuning prior to acceleration.
Commissioning with low-energy muon sources shows the system's capability to identify low-energy muon signals and measure beam profiles.
\end{abstract}


\maketitle


\section{Introduction}\label{sec:intro}
\begin{figure*}[t]
    \centering
    \includegraphics[width=0.9\linewidth]{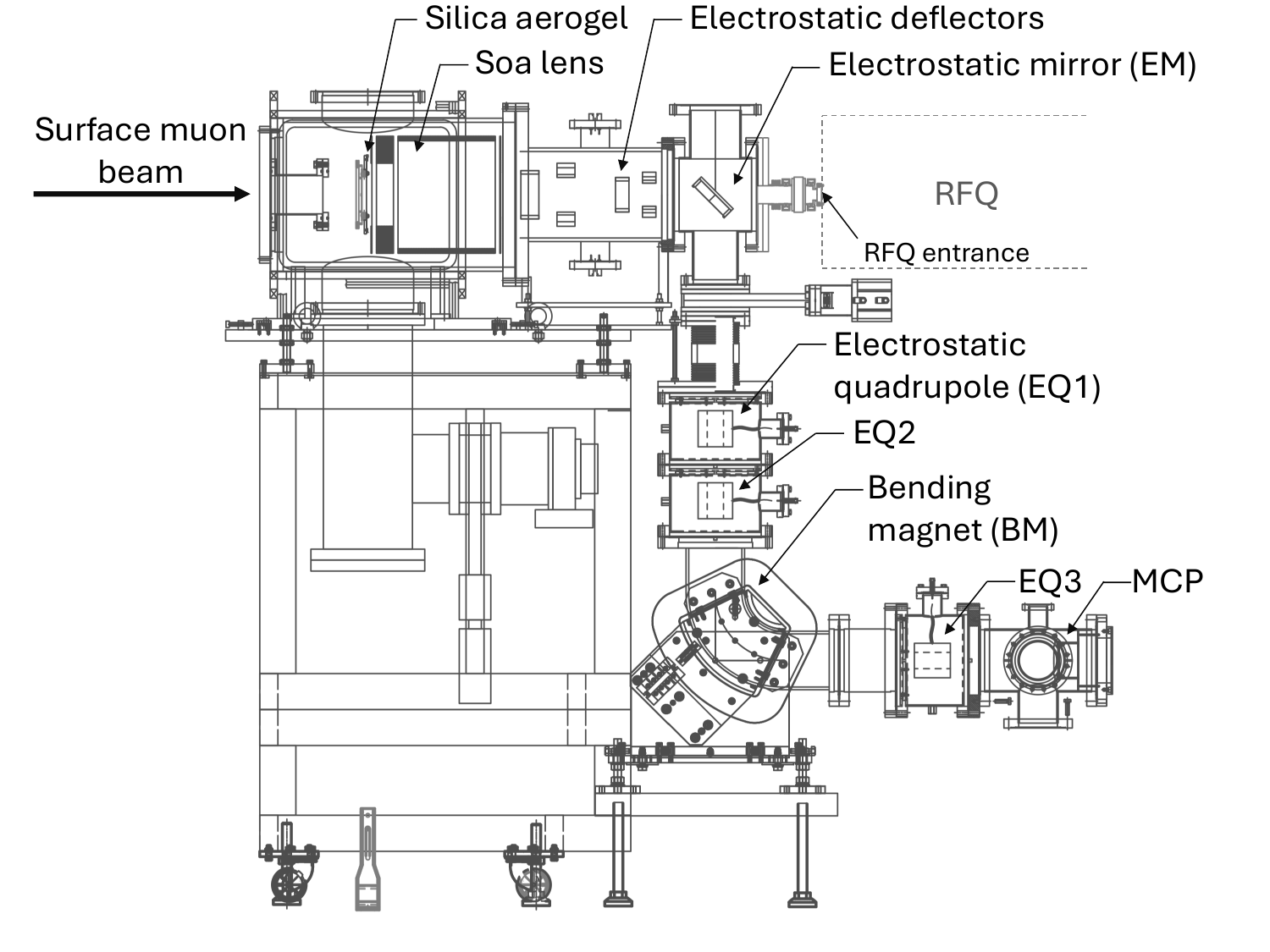}
    \caption{
    Side view of beamline including the diagnostic system. 
    In the upstream section, surface muons stop in the silica aerogel target and form muonium atoms, which are subsequently ionized by laser irradiation to produce low-energy muons. Generated muons are accelerated to \SI{5.7}{\kilo\electronvolt} and focused by the Soa lens, and transported toward the RFQ through the electrostatic deflectors. The diagnostic system, installed downstream of the deflectors, consists of EM, BM, EQ1--EQ3, and MCP.
    The low-energy muons are transported through EM, BM, and finally detected by the MCP. 
    EQ1 and EQ2 provide beam focusing, and EQ3 is used for the Q-scan. 
    The total path length from the silica aerogel to the MCP is approximately 2.5~m.
    }
    \label{fig:UDL}
\end{figure*}

Development of a low-emittance muon beam is an important step toward next-generation muon science. 
Such a beam enables high-precision measurements of fundamental physics parameters, such as the muon magnetic dipole moment, and the search for the electric dipole moment~\cite{abe2019new}. 
It is also expected to play key roles in future applications including muon microscopy for probing internal structures of materials~\cite{nagatani2023transmission} and muon colliders for probing physics beyond the Standard Model~\cite{hamada2022mutristan,boscolo2019future,boscolo2018low}.
Conventional high-intensity muon beams are predominantly produced from the decay of pions generated in proton-irradiated targets.
Such a muon beam is a tertiary beam and has a large phase space volume ($\mathcal{O}(10^{3})$ $\pi$ $\mathrm{mm}\cdot\mathrm{mrad}$), which leads to poor matching to the acceptance of the accelerator ($\mathcal{O}(1)$ $\pi$ $\mathrm{mm}\cdot\mathrm{mrad}$) and consequently low acceleration efficiency. 
A promising approach to overcome this issue is to cool the muon beam and subsequently accelerate it~\cite{neuffer1993muon,aritome2025acceleration}. 
One such positive muon acceleration scheme, proposed in Ref.~\cite{aritome2025acceleration}, transports a high-intensity surface muon beam~($\mu^+$) to an aerogel target placed in a vacuum chamber, where a fraction of the surface muons stop ~\cite{beer2014enhancement}. 
The stopped muons capture electrons and form muonium atoms~($\mu^+e^-$). A fraction of the muonium atoms is emitted from the downstream surface of the aerogel into the vacuum with a kinetic energy of approximately \SI{25}{\milli\electronvolt}. 
Laser ionization of the muonium releases the bound electron, leaving the positive muon~\cite{miyake2002ultrasensitive}.
These muons are extracted and accelerated to \SI{5.7}{\kilo\electronvolt}, focused by electrostatic lenses, and steered by electrostatic deflectors before being injected into a radio-frequency quadrupole (RFQ) accelerator.
The beam charge of the muon beam at the RFQ entrance is on the order of \si{\femto\coulomb}.

The RFQ requires the injection of the muon beam whose parameters are matched to its acceptance to avoid beam loss and emittance growth.
In particular, the trajectory of the low-energy muon beam is readily perturbed by ambient magnetic fields due to its small momentum.
Proper injection into the RFQ requires monitoring of the beam centroid position to verify trajectory correction. 
In addition, efficient acceleration in the RFQ requires beam matching to the RFQ acceptance, for which accurate measurement of the beam emittance and Twiss parameters at the injection point is essential.
However, the low intensity and low energy of the muon beam, as well as background particles coming from the upstream muon transport line, make the accurate determination of beam parameters challenging.
Conventional beam diagnostic techniques~\cite{jones2014introduction}, such as invasive profile monitors and non-destructive diagnostics based on beam current, are difficult to apply because the signal is too weak compared to the background.

In this work, we develop a diagnostic system to accurately evaluate the parameters of the low-energy and low-intensity muon beam while separating the signal from the background and maintaining compatibility with accelerator operation.
The system was installed between the muon source and the RFQ and is designed for flexible switching between acceleration and diagnostic modes.
Its performance is then demonstrated through beam-based measurements to verify its capability for signal identification and beam parameter measurements under realistic operating conditions.
The experiments were carried out at the Materials and Life Science Experimental Facility~(MLF) of the Japan Proton Accelerator Research Complex~(J-PARC).

This paper is organized as follows. 
Section~\ref{sec:design} describes the requirements and the system design. 
Section~\ref{sec:reconstruction} describes how the Twiss parameters at the RFQ entrance are reconstructed from the measured beam profiles and discusses the expected reconstruction accuracy. 
Section~\ref{sec:measurement} presents the measurements and evaluates the performance of the system.
Section~\ref{sec:conclusion} summarizes the conclusions.

\section{Diagnostic System Design and Specifications \label{sec:design}}
The diagnostic system was designed to fit within the limited available space upstream of the RFQ and to enable measurement of the beam parameters relevant to RFQ matching.
Since accurate beam parameter measurements are required to limit emittance growth in the RFQ, the requirements for the diagnostic system are as follows:

\begin{enumerate}
    \item Easy switching of the beam transport path between the acceleration mode, in which the beam is injected into the RFQ, and the diagnostic mode, in which the beam is directed to the diagnostic system, without interrupting accelerator operation.
    \item A compact configuration suitable for installation in the limited space between the muon cooling section and the RFQ.
    \item Background suppression through energy and momentum filtering.
    \item Measurement of beam parameters, such as the beam emittance and Twiss parameters, with an accuracy of approximately 10\% to limit the emittance growth within 10\%.
\end{enumerate}

\begin{figure}[t]
    \centering
    \includegraphics[width=\linewidth, 
    trim=2.5cm 8.9cm 7cm 2.7cm, clip]{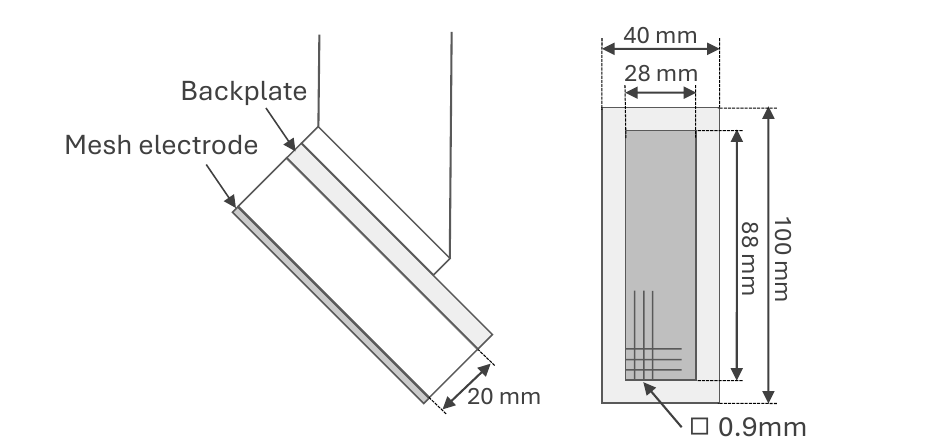}
    \caption{
    Schematic of the electrostatic mirror (EM). 
    The left panel shows a side view of the EM, consisting of a positively biased flat backplate and a grounded mesh electrode. 
    The electrodes are separated by \SI{20}{\milli\meter} and tilted by 45$^\circ$ with respect to the beam axis. 
    The right panel shows the front view of the electrode geometry. 
    The backplate has dimensions of \SI{40}{\milli\meter} in width and \SI{100}{\milli\meter} in height, while the mesh electrode has dimensions of \SI{28}{\milli\meter} in width and \SI{88}{\milli\meter} in height. 
    The mesh has an aperture of \SI{0.9}{\milli\meter}.
    }
    \label{fig:EM}
\end{figure}

\begin{figure}[t]
    \centering
    \includegraphics[width=1\linewidth]{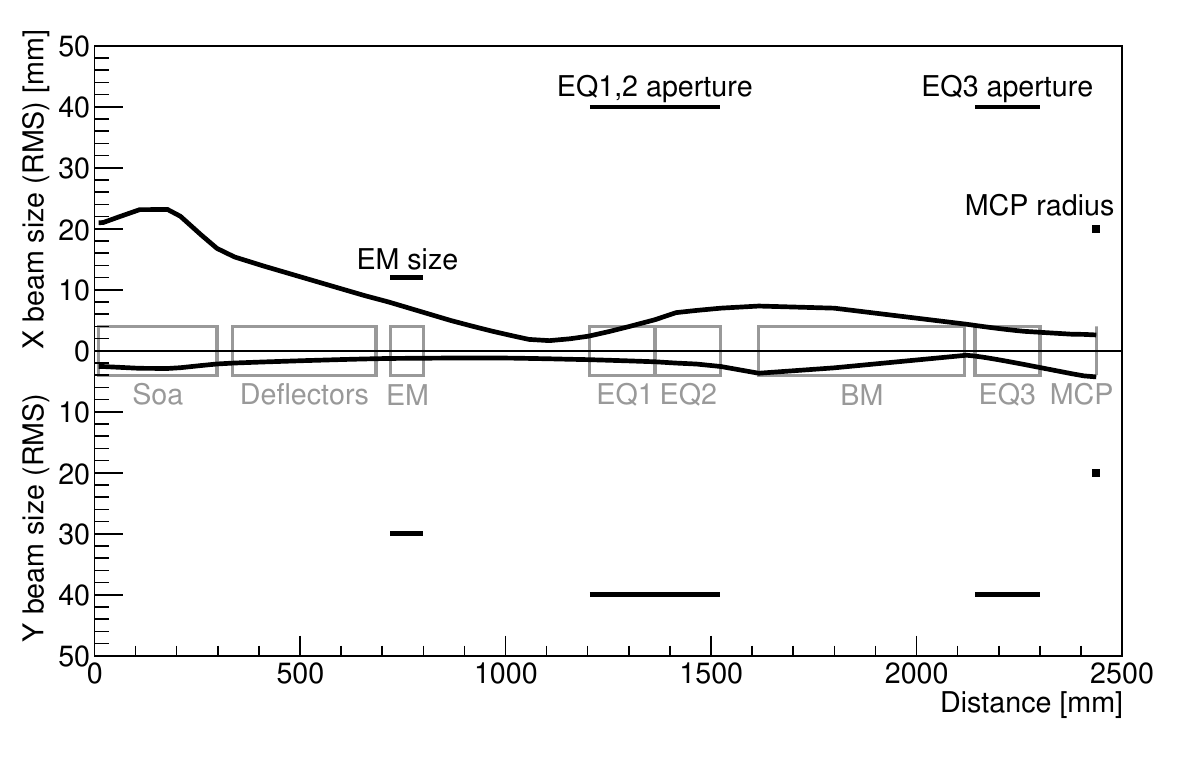}
    \caption{
    Beam envelope as a function of the distance from the silica aerogel. 
    The curves in the upper and lower panels are the horizontal ($x$) and vertical ($y$) RMS beam sizes, respectively. 
   The horizontal lines indicate the aperture sizes of the beamline components (the EQs and the MCP). The half-widths of the BM duct (\qty{37}{\milli\meter} in x and \qty{59}{\milli\meter} in y) are not shown.
    }
    \label{fig:beamenvelope}
\end{figure}

Figure~\ref{fig:UDL} shows the schematic of the beamline including the diagnostic system.
The upstream section consists of components for muon production, beam transport, and beam steering toward the RFQ.
The ambient magnetic field is continuously monitored using ten magnetic field probes installed around the target and the beamline.
The field cancellation coils are installed so as to cover the region from the silica aerogel target to the Soa lens~\cite{soa1959systematische}. 
Four electrostatic deflectors are installed between the Soa lens and the RFQ. The two upstream deflectors have an effective diameter of \SI{12}{\centi\meter}, whereas the two downstream deflectors have an effective diameter of \SI{8}{\centi\meter}. These deflectors are used for fine adjustment of the beam trajectory in the horizontal and vertical directions.
Following the electrostatic deflectors, we employ an electrostatic mirror (EM)~\cite{janka2024improving} to vertically deflect the beam towards the diagnostic section. 
EM, used for energy selection, can be removed from the beam path by a retraction mechanism so that the beam passes to the downstream accelerator. 
EM consists of a positively biased flat backplate and a grounded mesh electrode. The two electrodes face each other, are separated by \SI{20}{\milli\meter}, and are tilted by \SI{45}{\degree} with respect to the beam axis as shown in Fig.~\ref{fig:EM}. 
The backplate has dimensions of \SI{40}{\milli\meter} in width and \SI{100}{\milli\meter} in height.
The mesh electrode has dimensions of \SI{28}{\milli\meter} in width and \SI{88}{\milli\meter} in height and is made of SUS316L stainless steel with a thickness of \SI{0.1}{\milli\meter}. The mesh has an aperture of \SI{0.9}{\milli\meter} and a pitch of \SI{1.0}{\milli\meter} to maintain field uniformity while permitting beam transmission.
A bias voltage of \SI{5.7}{\kilo\volt}, corresponding to the extraction energy of the muon beam, is applied to the backplate.
A dipole magnet (bending magnet, BM) further deflects the beam by \SI{90}{\degree} and is used for momentum selection.
The magnet has a bending radius of \SI{150}{\milli\meter}, is air-cooled and operates at a central magnetic field of \SI{2.63e-2}{\tesla}. 
The combination of EM and BM rejects background particles from reaching a microchannel plate (MCP) located at the end of the system. 
In addition, there are three electrostatic quadrupoles. 
Each quadrupole has an aperture diameter of \SI{80}{\milli\meter} and an electrode length of \SI{80}{\milli\meter}. 
To suppress fringe field effects and improve field uniformity, donut-shaped end guards are attached to the electrode ends. 
Two of them (EQ1 and EQ2) are installed between EM and BM and provide beam focusing in combination with the edge focusing of BM.
In the present configuration, EQ1 is operated at \SI{0}{\kilo\volt}, while EQ2 is set to \SI{0.1}{\kilo\volt} to provide horizontal focusing and vertical defocusing. 
The last one (EQ3) is installed immediately upstream of the MCP to perform a quadrupole scan (Q-scan)~\cite{minty2003measurement} for beam parameter measurement.
The beam envelope throughout the system is shown in Fig.~\ref{fig:beamenvelope}.

Two types of MCP are used to detect muons, depending on the parameters to be measured.
For the measurement of the beam intensity and its time structure, a single-anode MCP (SA-MCP; Hamamatsu F9892-21) is used which has an effective diameter of \SI{42}{\milli\meter} and an open-aperture ratio of 60\%~\cite{hamamatsu_mcp_2025}.
The output waveform is sampled by a digitizer (CAEN, V1751) with a sampling rate of \SI{1}{\giga S \per \second}.
For the transverse profile measurements, an MCP coupled with a phosphor screen (BPM-MCP; Hamamatsu F2225-21P-Y003)~\cite{kim2018development} is used.
The effective detection area is \SI{40}{\milli\meter} in diameter.
The image of the scintillation light emitted from the phosphor screen (P47) is captured by a CCD camera (PCO pco.1600) with a timing gate set to \SI{500}{\nano\second}.

\section{Reconstruction of beam parameters}\label{sec:reconstruction}
The transverse beam properties relevant for beam matching are characterized by the emittance and the Twiss parameters. 
The beam parameters at the RFQ entrance are inferred from those measured in the diagnostic system through beam transport analysis.
Since direct measurement at the RFQ entrance is not practical in the present configuration, this section describes the reconstruction procedure used to estimate the Twiss parameters at the RFQ entrance from measurements performed in the diagnostic system.
For a Gaussian beam transported through a linear optical system, the phase space is related by the transfer matrix,
\[
\mathbf{x}_{\mathrm{RFQ\ entrance}} = R \,\mathbf{x}_{\mathrm{EQ3\ entrance}} \, ,
\]
where $\mathbf{x}_{\mathrm{RFQ\ entrance}} = (x,x')^T$  and  $\mathbf{x}_{\mathrm{EQ3\ entrance}}$ are the transverse phase space vectors at the RFQ entrance and the EQ3 entrance, respectively. $R$ denotes the transfer matrix of the diagnostic system.
This relation establishes a one-to-one correspondence between the beam parameters at the EQ3 entrance and those at the RFQ entrance.

To verify this correspondence, we constructed a particle tracking simulation including the electromagnetic field distribution of each beamline component and compared the result with that calculated by the transfer matrix.
The transfer-matrix calculation also takes into account the higher-order field components of the electric and magnetic fields.
If the two approaches agree, the one-to-one correspondence is validated and the beam distribution at the RFQ entrance can be determined by using the inverse transfer matrix $R^{-1}$.
The simulation is based on musrSim~\cite{sedlak2012musrsim}, a simulation framework based on Geant4~\cite{collaboration2003geant4} and ROOT~\cite{brun1997root}. 
Since the initial spatial extent of the low-energy muon beam is determined mainly by the laser irradiation region in the muonium source, we tested several plausible input beam distributions corresponding to experimentally accessible laser spot sizes.
These included a symmetric Gaussian distribution with identical transverse widths in the x and y directions, as well as an asymmetric distribution with a horizontal beam size larger than the vertical one, reflecting the elongated ionization region along the laser path.
To quantify the agreement between the two RMS ellipses, the mismatch factor is used. Following Ref.~\cite{Wangler2008}, the mismatch factor ($M$) is defined as
\begin{align}
    M &= \left[ 1 + \frac{\Delta + \sqrt{\Delta(\Delta + 4)}}{2} \right]^{1/2} - 1, \\
    \Delta &= (\Delta \tilde{\alpha})^{2} - \Delta \tilde{\beta} \, \Delta \tilde{\gamma}.
\end{align}
where $\Delta \tilde{\alpha}$, $\Delta \tilde{\beta}$, and $\Delta \tilde{\gamma}$ denote the differences in the emittance-normalized Twiss parameters between the two RMS ellipses.
A mismatch factor of zero corresponds to perfect agreement of the Twiss parameters.

For all the tested input beam distributions, both the relative difference in emittance between the two ellipses and the corresponding mismatch factor are $O(10\%)$.
This indicates that the ellipse represented by the Twiss parameters is well reproduced.
These results support the one-to-one correspondence between the transverse phase space coordinates at the RFQ entrance and at the EQ3 entrance through the transfer matrix. 
Accordingly, the beam parameters at the RFQ entrance can be reconstructed with the required accuracy by applying the inverse transfer matrix $R^{-1}$ to the beam parameters measured at the EQ3 entrance, thereby enabling evaluation of the RFQ matching condition.
The remaining discrepancies within $O(10\%)$ originate mainly from nonlinear effects such as electric fields in EM and EQs. 

\section{Evaluation of the diagnostic system}\label{sec:measurement}
The diagnostic system was developed for the H2 area at the MLF  where a low-energy muon source described in Ref.~\cite{abe2019new} is being commissioned.
Since stable beam operation in the H2 area was not yet available during this stage, the system was evaluated using low-energy muon beams in two experimental areas.
The measurements in the H2 area were used to demonstrate signal identification after installation of the system, whereas the measurements in the S2 area were used to evaluate the reproducibility of the beam profile measurements and the response of the beam size to the EQ3 focusing strength (Section~\ref{subsec:stability} and Section~\ref{subsec:qscan}).
The S2 area provides stable low-energy muons produced using the same laser-ionization scheme as reported in Ref.~\cite{aritome2025acceleration}.
These evaluations were conducted during dedicated beam time from October to December 2025.
This section presents the results of the evaluations.

\subsection{Identification of Muons}\label{subsec:identification}
\begin{figure}[t]
    \centering
    \includegraphics[width=1\linewidth]{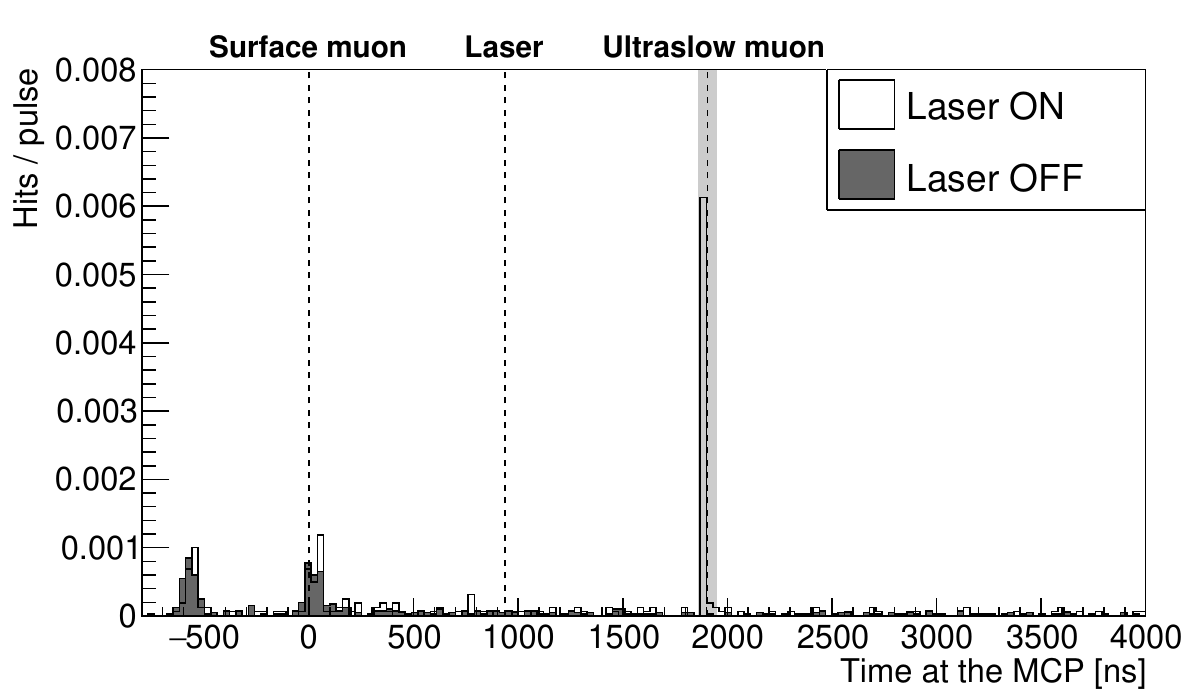}
    \caption{
        Time distributions measured with (open) and without (filled) the ionization laser. 
        The time origin ($t=$\SI{0}{\nano\second}) is defined as the mean of the arrival times of the two surface-muon bunches at the aerogel target. The vertical dashed lines correspond to the time origin, the laser-irradiation time, and the expected arrival time of the low-energy muon signal.
        Peaks observed around \SI{-570}{\nano\second} and \SI{30}{\nano\second} under both laser-on and laser-off conditions are attributed to  signals originating from positrons produced upstream of the beamline.
        A clear peak is observed around \SI{1900}{\nano\second} under the laser-on condition, corresponding to low-energy muons produced by laser ionization and transported to the MCP. 
        No significant signal is observed in the same time region without laser.
    }
    \label{fig:h2_tof}
\end{figure}

Figure~\ref{fig:h2_tof} shows the time distribution of hits detected by the MCP in the H2 area with and without the ionization laser.
The surface muon beam consists of two bunches separated by \protect\SI{600}{\nano\second}, and the time origin ($t=$\SI{0}{\nano\second}) is defined as the mean of their arrival times at the aerogel target.
In the H2 area, the surface muon beam has a pulsed time structure, and the laser irradiation was performed after the second muon pulse.
Under the laser-on condition, a clear peak is observed at \SI{1900}{\nano\second} which is consistent with the time of flight expected from the simulation. 
In contrast, no statistically significant events are observed in the same time region under the laser-off condition. 
These results confirm that the delayed hits around \SI{1900}{\nano\second} originate from low-energy muons produced by the laser ionization of muonium.
The signal-to-background ratio in the signal time window (\SI{1855}-\SI{1945}{\nano\second}) is approximately 60, demonstrating that low-energy muons can be clearly identified with this system.

\subsection{Measurement reproducibility\label{subsec:stability}}
\begin{figure}[t]
    \centering
    \includegraphics[
        width=0.95\linewidth,
        trim=0cm 0cm 10cm 0cm,
        clip
    ]{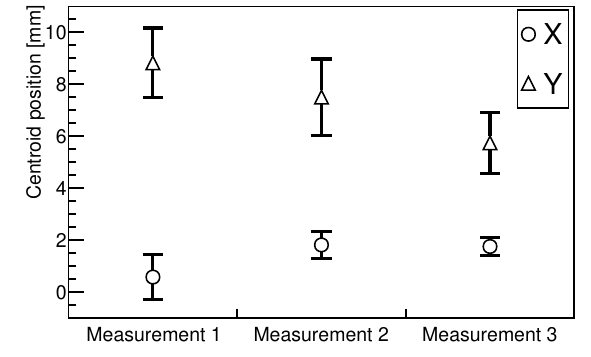}
    \vspace{8mm}
    \includegraphics[
        width=0.95\linewidth,
        trim=10cm 0cm 0cm 0cm,
        clip
    ]{reproducibility.pdf}
    \caption{
        Beam centroid positions and RMS beam sizes measured in three runs for stability evaluation of the diagnostic system.
        The quoted uncertainties were obtained by adding the statistical and systematic uncertainties in quadrature.
    }
    \label{fig:reproducibiliry}
\end{figure}

For an online diagnostic system used for beam matching, the measured beam centroid position and RMS beam size must be reproducible under nominally identical beam conditions, because these quantities are used to evaluate the beam trajectory and transverse beam parameters.
The measurement reproducibility of the diagnostic system was evaluated by repeating beam profile measurements in three independent runs performed within approximately one day under nominally identical beam conditions in the S2 area. 
Figure~\ref{fig:reproducibiliry} summarizes the measured beam centroid positions and RMS beam sizes obtained from these measurements. 
The measured beam parameters are consistent within the uncertainties. 

The quoted uncertainties were obtained by adding the
statistical and systematic uncertainties in quadrature.
The systematic uncertainty mainly originates from the background subtraction in the beam profile analysis.
Residual background contamination can distort the beam profile and lead to an incorrect evaluation of the beam centroid position and RMS beam size.
This systematic uncertainty was estimated from toy simulations by varying the subtracted background level within its statistical uncertainty and taking the resulting variations in the beam centroid position and RMS beam size.

\subsection{Beam profile measurement}\label{subsec:qscan}
\begin{figure*}[t]
    \centering
    \includegraphics[width=\linewidth]
    {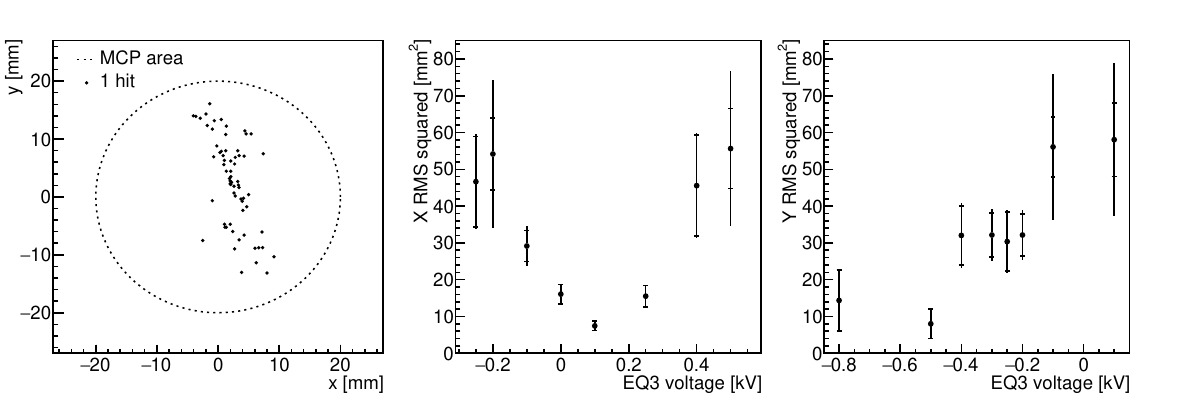}
    \caption{
    Beam profile measurement using the BPM-MCP.
    Left: A transverse beam profile obtained at the point of minimum horizontal beam size. Each point represents a single detected hit. The dashed circle corresponds to the sensitive area of the MCP.
    Center and right: The squared RMS beam sizes in the horizontal (center) and vertical (right) directions as functions of the EQ3 voltage.
    The inner (outer) error bars indicate the statistical (total) uncertainties.
    }
    \label{fig:profile}
\end{figure*}

The transverse beam profile was measured using the BPM-MCP.
Figure~\ref{fig:profile} shows the beam profile obtained at the EQ3 voltage that provides the strongest horizontal focusing.
Each point represents an individual detected hit, and the circle denotes the sensitive area of the MCP.

To investigate the dependence of the beam size on the quadrupole focusing strength, the voltage applied to EQ3 was scanned, and the beam profiles were measured at each point.
The resulting squared RMS beam sizes in the horizontal ($x$) and vertical ($y$) directions are shown in Fig.~\ref{fig:profile}.
The uncertainties were evaluated using the same procedure as that described in Sec.~\ref{subsec:stability}.
In both directions, the beam size varies systematically with the applied voltage, exhibiting the expected focusing and defocusing behavior of a quadrupole lens.

These results demonstrate that the diagnostic system is capable of measuring transverse beam profiles and quantitatively determining the beam size over a range of quadrupole strengths, which is necessary for the Q-scan measurement.
Determination of the Twiss parameters requires corrections to the measured beam sizes to account for the limited sensitive area of the BPM-MCP, as well as an accurate description of the beamline optics.
These studies will be performed in future work using a
larger data sample.

\section{Conclusion}\label{sec:conclusion}
A compact diagnostic system for low-energy muon beams before RF acceleration was developed.
The diagnostic system is designed to evaluate beam parameters at the entrance of the RFQ with sufficient accuracy. 
The measurements verified that the low-energy muon signal could be identified from the time distribution with a signal-to-background ratio of about 60.
Repeated measurements confirmed the reproducibility of the beam profiles.
Beam profiles were measured while scanning the EQ3 voltage. 
The observed variation in the beam size confirms the focusing and defocusing effects of EQ3.
This system enables evaluation of beam parameters for low-energy and low-intensity particle beams and will support the realization of low-emittance muon beams.

\begin{acknowledgments}
The authors would like to thank the J-PARC muon section staff for their support in the conduct of the experiment at J-PARC MUSE. The experiment was performed at the Materials and Life Science Experimental Facility of the J-PARC under a user program 2011MS06. We sincerely thank the Mechanical and Engineering Center of KEK for supporting the design and manufacturing of instruments. This work was supported in part by JSPS Kakenhi Grants No. 20H05625, 22K21350, 
24H00023, 
24K03211, 
25K24694,
MEXT Q-LEAP JPMXS0118069021,
and JST K-program JPMJKP24J4.
\end{acknowledgments}

\bibliography{main}
\end{document}